# Reconfigurable Current Distributions in Percolating Au Films and $CrO_2$ Powder Compacts

## *Beyond Geometrical Connectivity*

**E. Yu. Beliayev, I. G. Mirzoiev, V. A. Horielyi**

*B. Verkin Institute for Low Temperature Physics and Engineering of the National Academy of Sciences of Ukraine, Kharkiv 61103, Ukraine*

*Corresponding author: beliayev@ilt.kharkov.ua; beliayev@gmail.com

## Abstract

Despite fundamental differences in morphology, dimensionality, and microscopic transport, semicontinuous Au films and compacted $CrO_2$ powders converge on a common network-level principle: the electrically relevant network is a condition-dependent subset of the physical contact network. In Au films near the geometrical percolation threshold, increasing bias reversibly shifts the low-temperature response from activated nearest-neighbour hopping with negative magnetoresistance to a weak-localization-like regime with positive magnetoresistance, consistent with broader electronic participation of the fixed island network. In $CrO_2$ powder compacts, magnetic field mainly changes the spin-dependent resistances of existing intergranular junctions and thereby reorders competing current paths; temperature and measuring current can additionally expand or deplete the active network. Cohn's theorem provides a compact sensitivity principle for both cases: local resistance changes influence the measured response in proportion to the current carried by the affected links. The comparison identifies two limiting but interconvertible modes of reconfiguration - link enabling and link reweighting - and shows why geometrical connectivity alone is insufficient for interpreting transport in strongly inhomogeneous conductors.

## 1. Introduction

Percolation is usually introduced as a geometrical problem: conducting regions acquire macroscopic connectivity only when their concentration exceeds a critical value. That description is indispensable, but it is not sufficient for real disordered conductors. Once a geometrically continuous network exists, the resistances of its links may still span many orders of magnitude. The measured resistance is then controlled not by the entire connected structure, but by a smaller and condition-dependent set of current-carrying links.

This distinction is especially important when charge transfer between neighbouring conducting regions occurs by tunnelling or thermally assisted hopping. Exponential sensitivity to barrier width, barrier height, and activation energy produces a broad distribution of contact resistances. Cooling or reducing the applied bias suppresses the least favourable contacts and can concentrate the current in only a few paths. Conversely, an external perturbation may make additional contacts accessible or change the relative ranking of already active links without creating or destroying physical contacts.

The comparison developed here deliberately uses two systems that are very different at the microscopic level. Semicontinuous Au films are two-dimensional island networks close to a geometrical percolation threshold, whereas compacted $CrO_2$ powders are three-dimensional assemblies of ferromagnetic particles separated by dielectric surface layers. Their dissimilarity is useful: any common conclusion cannot be attributed to a shared morphology, dimensionality, or microscopic conduction mechanism.

In the Au films, applied bias reversibly changed the resistance by several orders of magnitude and transformed the observed low-temperature response from nearest-neighbour hopping to a weak-localization-like regime [1]. Related films displayed simultaneous one- and two-dimensional quantum-interference signatures [2], while mesoscopic conductance fluctuations revealed extreme sensitivity to a small number of unavoidable high-resistance bridges [3] and [4]. These results point to a current distribution that narrows or broadens within an essentially fixed island structure.

$CrO_2$ powder compacts provide a complementary magnetic case. Their intergranular conductance depends on both the tunnel barrier and the relative directions of neighbouring magnetic moments [5]. With cooling, the active network becomes increasingly selective, so the field of a magnetoresistance maximum can depart markedly from the coercive field obtained from bulk magnetization [6], [7], and [8]. Additional branch crossings and nonmonotonic magnetoresistance indicate history-dependent redistribution among a small number of transport paths [9].

The aim of this article is not to assign Au and $CrO_2$ to a common universality class or to reconstruct a unique spatial current map from macroscopic resistance measurements. Instead, we distinguish the physical contact network from the electrically accessible network and from the still smaller current-carrying backbone. Within this hierarchy, we identify two limiting modes of reconfiguration: link enabling, which changes the breadth of the active network, and link reweighting, which changes the relative importance of competing paths. The experimental sections then test these modes and show how a single system can move between them.

## 2. Transport Connectivity and Two Modes of Reconfiguration

Three levels of connectivity must be separated. The physical contact network comprises the islands, grains, constrictions, gaps, and interfaces created during sample preparation. The electrically accessible network is the subset of those contacts whose resistances are low enough to participate under the chosen temperature, bias, magnetic field, and history. The current-carrying backbone is narrower still: it contains the branches that carry a substantial fraction of the total current and the bottlenecks that cannot be bypassed without a large resistance penalty.

For a linear resistor network at fixed total current $I$, the sensitivity of the equivalent resistance $R$ to the resistance r of an individual link is given by Cohn's theorem [10]:

$$\partial R/\partial r_\alpha = (i_\alpha/I)^2 \quad (1)$$

Here i is the current through the selected link. The theorem expresses the central weighting principle: a local resistance change matters macroscopically only to the extent that the affected branch carries current. In a nonlinear network, Eq. (1) is not a global constitutive law; it applies only to a differential response after local linearization around a specified operating point. We therefore use it as a sensitivity principle, not as a microscopic model of field-assisted tunnelling or magnetic reversal.

The same principle also clarifies why bulk-averaged material parameters can be misleading. A large magnetic volume may contribute strongly to magnetization but weakly to resistance if it lies outside the dominant paths. Conversely, a small inter-island gap or a single magnetic junction can control the total response if it belongs to a critical branch. The relevant question is thus not only whether the sample is connected, but which connected elements are selected and weighted by the current.

*Table 1. Network-level contrast between the two experimental systems.*

| Feature | Semicontinuous Au films | $CrO_2$ powder compacts |
|---|---|---|
| **Physical structure** | Two-dimensional metallic islands and constrictions near a geometrical percolation threshold | Three-dimensional network of ferromagnetic particles separated by dielectric surface layers |
| **Main control variable** | Applied bias and temperature | Magnetic field, field history, measuring current, and temperature |
| **Dominant local change** | Reduced activation or tunnelling cost of high-resistance inter-island links | Change in spin-dependent resistance as neighbouring magnetic moments reorient |
| **Dominant network response** | Link enabling: expansion or contraction of the active network | Link reweighting: change in the ranking of competing current paths |
| **Characteristic signatures** | Activated-to-WL-like crossover, MR sign reversal, mixed 1D-2D quantum corrections | Transport-bulk coercivity mismatch, branch crossings, current and sweep-rate dependence |

Link enabling denotes a change in electronic accessibility. Contacts that were effectively inactive become competitive, so the active network gains parallel routes and the voltage drop becomes less concentrated at a few bottlenecks. Link reweighting denotes a change in the conductance ranking of links that are already part of the accessible network. It can redirect current between competing paths without a comparable increase in the number of usable contacts.

These modes are limiting descriptions rather than mutually exclusive phases. Enabling a new contact necessarily reweights the remaining paths, and a sufficiently large change in bond ranking can remove one path from practical participation while activating another. The distinction identifies the dominant action of the external control parameter. It also avoids equating transport dimensionality with sample geometry: a two-dimensional film can contain quasi-one-dimensional bottlenecks, while a three-dimensional compact can conduct through only a few chain-like routes.

## 3. Bias-Controlled Current Redistribution in Percolating Au Films

### 3.1 Bias-controlled crossover between transport regimes

The most pronounced bias dependence was observed in a semicontinuous Au film of nominal thickness about 3.25 nm, deposited on sapphire near 54 K [1]. The film consisted of metallic islands or larger percolation clusters separated by vacuum gaps and narrow constrictions. Its low-temperature current-voltage characteristics were strongly nonlinear, yet the transport state could be changed reversibly over the available bias range.

At $U \leq 0.05$ V and $T$ below about 20 K, the resistance followed an Arrhenius-like law

$$R(T) \propto \exp(T_0/T) \tag{2}$$

with $T_0 \approx 20$ K. Below 1 K, the voltage dependence was approximately

$$R(U) \propto \exp(U_0/U) \tag{3}$$

with $U_0 \approx 2$ V. The resistance reached about $10^7$ Ω, and the negative temperature coefficient identified a strongly localized inter-island transport regime [1]. The Arrhenius form is treated here as an empirical signature of nearest-neighbour hopping controlled by a restricted set of dominant inter-island barriers, rather than as evidence for a specific variable-range-hopping law. Strong-field hopping theory provides a relevant network-level precedent: an applied field can lower dominant hopping barriers and modify optimal percolation paths, producing a strongly non-Ohmic response [11]. The comparison is qualitative because the Au film exhibits nearest-neighbour rather than Efros-Shklovskii variable-range hopping.

At biases of about 5-11 V, the sheet resistance fell to approximately 5 kΩ and its temperature dependence became weak and roughly logarithmic over 5-55 K. The magnetoresistance was then consistent with quantum weak-localization corrections in a disordered two-dimensional metal [1] and [12]. The data therefore establish a bias-controlled change in the dominant transport response: activated inter-island transfer controls the low-bias state, whereas the high-bias state increasingly samples metallic regions that support phase-coherent diffusion.

This result should not be interpreted as proof of an equilibrium thermodynamic metal-insulator transition. In a percolating film, a large fraction of the applied voltage can drop across only a few critical junctions, so local electric fields and local dissipation are highly nonuniform. Field-assisted barrier lowering, suppression of activation, electronic overheating, and redistribution of the voltage drops may all contribute. The experimentally secure conclusion is narrower but still significant: applied bias changes which parts of the fixed island network participate effectively in conduction.

In network language, Au is an enabling-dominated case. As the most resistive inter-island links become less costly, additional routes carry appreciable current. The active network broadens, the resistance becomes less dominated by individual bottlenecks, and the response evolves toward that of the metallic regions embedded in the film.

### 3.2 Magnetoresistance, resistance hierarchy, and effective dimensionality

The reversal of the magnetoresistance sign provides an independent marker of this redistribution. At high bias, Au exhibits positive magnetoresistance associated with weak localization in the presence of strong spin-orbit scattering. At low bias, the magnetoresistance is negative and approximately follows

$$\Delta R(H)/R(0) \propto -H^2/T \tag{4}$$

under the same conditions in which the resistance is activated [1]. The low-bias negative magnetoresistance should not be identified simply with suppression of the high-bias weak-localization term. It occurs in a nearest-neighbour-hopping regime, and a study devoted specifically to this effect concluded that its microscopic origin remained unresolved [13]. We therefore use the sign change as an empirical indicator of a change in the dominant transport regime, not as evidence for a unique hopping-magnetoresistance mechanism.

The sign reversal did not occur at a universal sheet resistance. It was observed near 36 kΩ at 1.5 K, 16 kΩ at 3 K, and 5.7 kΩ at 15 K [1]. A single macroscopic resistance is therefore insufficient to locate a localization boundary in the heterogeneous film. The measured value combines metallic-region resistance, tunnel-junction resistance, and the geometry of the current distribution, all of which respond differently to temperature and bias.

Weak-localization fits also revealed a hierarchy of resistance scales. The effective sheet resistance associated with phase-coherent motion inside metallic islands was up to about an order of magnitude smaller than the directly measured sheet resistance of the complete film [1]. Although the full current-voltage characteristic is nonlinear, the response to a small magnetic-field-induced correction at fixed bias can be considered using the tangent network of differential resistances. Eq. (1) then describes only the local sensitivity around that operating point. A metallic region on a dominant path may carry $i_\alpha \approx I$, so its absolute weak-localization correction is not eliminated; its relative contribution to $\Delta R/R$ is diluted by the much larger, weakly field-insensitive inter-island resistance included in the total resistance. Contributions from metallic regions in weak-current branches are additionally suppressed by the factor $(i_\alpha/I)^2$.

A second Au film, of nominal thickness about 3.56 nm, provided complementary evidence that the active current distribution can have a lower effective dimensionality than the specimen [2]. Above roughly 3 K, the quantum corrections were predominantly two-dimensional. Below that temperature, an additional low-field magnetoresistance component and a deviation from the two-dimensional logarithmic temperature dependence were consistent with coexisting one- and two-dimensional weak-localization and electron-electron-interaction contributions.

The estimated channel dimensions in that analysis were model dependent because the local sheet resistance, diffusion coefficient, and phase-coherence length of the inferred constrictions were not known independently. They should not be read as a direct geometrical map of the current. The robust observation is the coexistence of dimensional signatures: broad metallic regions behave two-dimensionally, while narrow links within the same percolating structure contribute a quasi-one-dimensional response.

Mesoscopic conductance fluctuations complete this picture. Near the threshold, a wide film can be controlled by a small number of unavoidable bridges, so reproducible changes in individual bottlenecks become visible in the total conductance [3] and [4]. Cooling or lowering the bias narrows the active distribution and enhances this sensitivity; increasing bias reverses the trend by enabling additional links.

## 4. Magnetic Reweighting in $CrO_2$ Powder Compacts

### 4.1 Transport-selected magnetic response

Compacted $CrO_2$ powders form three-dimensional networks of ferromagnetic metallic particles separated by thin dielectric surface layers. Charge transfer occurs through a random array of magnetic tunnel junctions whose conductances depend on barrier properties and on the relative directions of neighbouring magnetic moments [5]. During a magnetic-field sweep, the mechanical contact network is essentially fixed; the field changes the magnetic configuration and therefore the conductance hierarchy of the junctions.

At elevated temperatures within the ferromagnetic state, many intergranular paths contribute to conduction, and magnetotransport averages over a substantial fraction of the particles. Cooling suppresses the less favourable junctions and concentrates current in fewer low-resistance chains. The electrically sampled magnetic ensemble then becomes smaller than the full magnetic volume, so transport and magnetization need not characterize the same reversal process.

This difference appears directly in the fields $H_p$ and $H_c$. $H_p$ is the field of a magnetoresistance maximum, whereas $H_c$ is the coercive field obtained from bulk magnetization. The two quantities approach one another when many paths contribute, but their ratio changes strongly and nonmonotonically on cooling [6], [7], and [8]. At intermediate temperatures $H_p$ often exceeds $H_c$, indicating that the dominant paths sample particles or aggregates with above-average coercivity. At the lowest temperatures $H_p$ may fall below $H_c$, consistent with a depleted network in which large multidomain particles with favourable contact areas and comparatively low domain-wall-mediated reversal fields become disproportionately important.

These microscopic assignments remain interpretive rather than unique. The firm conclusion is that $H_p$ is a transport-weighted coercive scale, not a modified bulk coercive field. Eq. (1) explains why: a resistance change at a junction carrying a large fraction of the current can dominate the measured loop even when the associated particles represent only a small fraction of the magnetic volume.

### 4.2 Field-, current-, and time-dependent reconfiguration

Magnetic field changes the relative magnetization directions across individual tunnel junctions. In single-domain particles, reversal proceeds through rotation or switching of the net magnetic moment; in multidomain particles, it can involve domain-wall motion and redistribution of domain volumes. Because these processes occur at different fields and with different histories, they change the spin-dependent resistance ranking of competing current paths.

If two routes have comparable total resistance, a modest change at one critical junction can transfer a substantial fraction of current from one route to the other. Additional crossings between increasing- and decreasing-field branches of the magnetoresistance loops are consistent with such history-dependent reranking [9]. The word switching here denotes a change in the dominant path; the smooth published curves do not establish that the transfer is microscopically abrupt or bistable.

This path dependence also explains why resistance cannot in general be represented as a unique function of bulk magnetization. States with similar total magnetization can have different resistances if the critical contacts have different local configurations, while magnetic reversal outside the active network can change the bulk moment with little transport consequence.

The measuring current adds a second control channel. As the network depletes on cooling, current density and voltage drop rise at the critical contacts. Local Joule heating, barrier lowering, and activation of additional junctions can then broaden the active network. Measurements showed a strong and temperature-dependent $H_p(I)$ relation [8]. Near 10 K, increasing current initially reduced $H_p$, consistent with overheating or voltage-assisted enabling of additional links. At lower temperatures $H_p$ increased with current; spin-polarized-current effects on multidomain particles were proposed, but heating, barrier modification, and current redistribution could not be separated uniquely.

This current dependence is important conceptually because it links the two limiting modes. Magnetic field makes $CrO_2$ primarily a reweighting-dominated system, but sufficiently large bias can introduce link enabling by lowering or thermally bypassing the most resistive barriers. The observed tendency of transport-derived fields to move toward the bulk coercive scale as the active network broadens is consistent with this crossover [8].

The current distribution can also relax on experimentally accessible time scales. In Fe-containing $CrO_2$ powders, the magnetoresistance shape depended on magnetic-field sweep rate, and the anomalous high-field structure weakened as the observation time increased [14]. A slowly relaxing magnetic configuration can strongly affect resistance when it belongs to a critical junction even though its contribution to bulk magnetization is small. Sparse transport therefore acts as an amplifier of local magnetic dynamics.

## 5. From Limiting Modes to Mixed Reconfiguration

Link enabling and link reweighting describe different actions on the same underlying object: the distribution of bond conductances. Enabling lowers the effective cost of links in the high-resistance tail and increases the number of parallel routes that carry appreciable current. Reweighting changes the ordering among already competitive links and redirects current without necessarily broadening the network.

The Au films are enabling dominated because bias weakens activation and tunnel bottlenecks, revealing a progressively broader metallic network. The $CrO_2$ compacts are reweighting dominated with respect to magnetic field because field history changes the spin-dependent ranking of existing junctions. Yet the current-dependent $CrO_2$ results show that the modes are not fixed material labels: increasing bias can drive the same compact toward a mixed regime in which additional paths are enabled while the magnetic field continues to reweight them.

Nor is the distinction equivalent to smooth versus discontinuous response. A broad network can reconfigure continuously as many bond conductances evolve, whereas nearly degenerate sparse paths can exchange dominance over a narrow control interval. Whether that exchange appears as a smooth crossover, enhanced noise, or a resolved jump depends on topology, conductance correlations, hysteresis, and experimental bandwidth. The present data support path reranking but do not justify a universal bistability criterion or a threshold based only on the ratio between the largest and smallest bond resistances.

The two modes instead suggest experimentally testable diagnostics. Enabling should weaken activated temperature dependence, reduce the fraction of voltage dropped at individual bottlenecks, and make the response more representative of the extended network. Reweighting should produce stronger history dependence, branch-specific anomalies, and disagreement between transport-derived and bulk-averaged characteristic fields. Both effects should be strongest when cooling or low bias restricts conduction to only a few paths.

These diagnostics are already visible in the systems considered here. In Au, lowering bias or temperature strengthens nonlinearity, hopping signatures, mesoscopic sensitivity, and quasi-one-dimensional contributions. In $CrO_2$, lowering temperature enhances the mismatch between transport-derived and bulk coercive fields and strengthens path-specific hysteresis, whereas increasing measuring current can broaden the active network and reduce the distinction between transport and bulk magnetic scales.

Taken together, these diagnostics turn the distinction into more than a descriptive vocabulary. They identify whether an external parameter mainly changes the breadth of the active network or the ordering of its dominant links, and they show how the same material can cross between the two limits. The framework also explains why bulk-averaged parameters fail most conspicuously near a sparse backbone: the measured response is governed by the few links that carry the current, not by the majority of the sample.

## 6. Conclusions

Semicontinuous Au films and $CrO_2$ powder compacts demonstrate that the network relevant to transport is not identical to the physical contact network. In Au, increasing bias reversibly shifts transport from activated hopping to a weak-localization-like regime, consistent with broader electronic participation of inter-island links. In $CrO_2$, magnetic field primarily reweights spin-dependent tunnel junctions and redirects current among competing paths.

The two modes are not mutually exclusive. Current and local dissipation can enable additional links in $CrO_2$, while every newly enabled Au link also changes the weights of the pre-existing paths. Cohn's theorem, applied to the local differential network where required, supplies the common sensitivity principle: local changes become macroscopically important when they occur on branches carrying a large fraction of the current.

Spatially resolved current mapping, nonlocal transport, time-resolved noise, and systematic bias- and sweep-rate-dependent measurements provide direct next tests of this framework. In particular, they could distinguish continuous redistribution from intermittent changes between nearly degenerate paths and determine when a reweighting-dominated network crosses into a mixed or enabling-dominated regime.

The comparison therefore moves the interpretation of percolative transport beyond geometrical connectivity. Resistance, magnetoresistance, and transport-derived characteristic fields belong to a condition-dependent current distribution selected by temperature, bias, magnetic field, history, and relaxation time. Recognizing this selection provides a compact way to organize otherwise disparate anomalies without claiming a universal microscopic mechanism.